# Promises and Challenges in Continuous Tracking Utilizing Amino Acids in Skin Secretions for Active Multi-Factor Biometric Authentication for Cybersecurity

Juliana Agudelo,[a]   Vladimir Privman,[b]   Jan Halámek*[a]

[a] *Department of Chemistry, University at Albany, State University of New York, Albany, NY 12222 (USA), E-mail: jhalamek@albany.edu*

[b] *Department of Physics, Clarkson University, Potsdam, NY 13699 (USA)*

We consider a new concept of biometric-based cybersecurity systems for active authentication by continuous tracking, which utilizes biochemical processing of metabolites present in skin secretions. Skin secretions contain a large number of metabolites and small molecules that can be targeted for analysis. Here we argue that amino acids found in sweat can be exploited for the establishment of an amino acid profile capable of identifying an individual user of a mobile or wearable device. Individual and combinations of amino acids processed by biocatalytic cascades yield physical (optical or electronic) signals, providing a time-series of several outputs that, in their entirety, should suffice to authenticate a specific user based on standard statistical criteria. Initial results, motivated by biometrics, indicate that single amino acid levels can provide analog signals that vary according to the individual donor, albeit with limited resolution versus noise. However, some such assays offer digital separation (into well-defined ranges of values) according to groups such as age, biological sex, race, and physiological state of the individual. Multi-input biocatalytic cascades that handle several amino acid signals to yield a single digital-type output, as well as continuous-tracking time-series data rather than a single-instance sample, should enable active authentication at the level of an individual.

# 1. INTRODUCTION

## 1.1. Overview

The security of electronic devices such as smartphones or smart watches, which are often constantly connected to applications involving sensitive and personal information, is based on reliable authentication of the actual user/owner of the particular device. However, no single authentication methodology is foolproof. While theoretically, only the owner should know the passcode — be it a phrase or a numerical combination — passwords can be duplicated, e.g., by spying on the owner while unlocking the device. Pattern-based authentication can also be bypassed. For example, fingerprint molds can be "lifted" from fingerprints left on various surfaces, which allows unauthorized users to "trick" a fingerprint reader.[1] Another aspect of the problem has been the reluctance of the users to input their password/passcode too frequently. Both of these issues have led to the advent of active authentication approaches.[2,3] These typically include trace histories[2,4] that can be based on web-use (such as browsing history, application usage), and/or digital (facial recognition or speech analysis), physical (user's gait, touch/swipe dynamics), or global/location tracking (following the user's routine, GPS) that uses data collection and their analyses to determine that the user is an authorized person, whose additional verification with password will otherwise be requested on each access to any sensitive application. Active authentication requires various continuous authentication and tracking methodologies[2-5] that potentially involve not only the recording of various "histories," but also continuous monitoring of the user's patterns of behavior.

This article reviews a new concept and also outlines initial results for utilizing forensic biometrics to develop a new (bio)chemical approach to data collection for continuous active authentication and trace-history information gathering. The methodology is not based on the electronic (digital) or pattern (physical/optical) inputs, but rather on biochemical inputs: metabolites in the user's skin secretions. The approach is autonomous and can be used by all individuals who own or have access to technology that holds personal information. Specifically, we address a *continuous and unobtrusive user authentication and physiological state monitoring* approach using metabolites secreted by the skin. The latter capability: to monitor the physiological state of the user, is unique to such a "biochemical" approach as compared to other "electronic" or "physical" continuous tracking methods, which are more useful for monitoring "lifestyle/habits" information. Therefore, besides authentication and user tracking, this methodology can potentially enable a more convenient use of electronic devices for persons with certain disabilities.

Analytes that can be used as "input signals" for monitoring, include, but are not limited to, various amino acids — which we focus on here — as well as pyruvate, and other metabolites. These compounds are substrates, intermediates, and products of many metabolic reactions and processes. As a result, humans secrete different levels of these substances through eccrine glands as components of sweat.[6] We survey studies[6-11] and illustrate results, originally motivated by forensic applications, which suggest that biochemical metabolic profiles, when recorded as a sufficiently detailed, continuously tracked time-series, are different from individual to individual. Larger, more noticeable differences are caused by variations in physical/physiological attributes[11] such as biological sex, ethnicity, and age, but there are smaller differences found between individuals even with similar physical attributes.

Sweat is a biological sample that is generally easy to collect in a non-invasive and unobtrusive manner by having a small sensor at typical point(s) of skin contact with a device, making it an ideal target for new biometric-based security approaches. Ultimately, the reviewed approach should enable the development of a new methodology to allow the transitioning from and/or supplementing passcode-, data-, physical- and image-based security with *biochemical* tracking that includes security measures based on the unique metabolite levels' trace histories of the device owner/user.

## 1.2. Statement of the Problem

As described in the introduction, most approaches and technologies for secure authentication of device users are not hack-proof.[1] Furthermore, some of them — notably, frequent inputting of a passcode — are not user-friendly. Therefore, a new/additional authentication methodology based on biochemical analytes as "signals" will be very useful. This methodology involves targeting analytes present in sweat as "chemical inputs" for unobtrusive, continuous tracking and active authentication. A challenge lies in identifying the multi-analyte sets to probe (measured signals) and establishing the feasibility of user authentication based on the collected data, as well as developing biomolecular processing of the collected data for the latter purposes.

The analytes, numbering $N$, provide a time series $x_{n=1,\ldots,N}(t_0 + k\tau)$ of chemical concentrations $x_{n=1,\ldots,N}$ probed at certain time intervals $\tau$, starting with the initial time $t_0$. Biochemical processing in a patch/sensor then converts this set of the input analytes $x_{n=1,\ldots,N}$ into a set of $S$ output chemical and ultimately physical (electronic, optical) signals, $y_{s=1,\ldots,S}$. Usually, due to the nature of such processing, the conversion is "for the same time step": The output signals $y_{s=1,\ldots,S}(t_0 + k\tau + t_g)$ are obtained from $x_{n=1,\ldots,N}(t_0 + k\tau)$ probed at each specific time step, $k = 0, 1, 2, \ldots$, with typically some delay "gate time" (measurement time) $t_g > 0$, usually smaller than or comparable to $\tau$, which is required for the biochemical processing of the inputs into the outputs.

The "vector" (of $S$ values) time series of the output signals can then be analyzed for authentication. The role of statistical analysis[2-5] is to establish that the signal time-series $y_{s=1,\ldots,S}(t_0 + k\tau + t_g)$ is sufficient for providing an ongoing user authentication/verification and tracking/physiological-state monitoring for $k > k_{\text{reg}}$, after the initial "registration/identification" over the first several time steps $k = 0, 1, \ldots, k_{\text{reg}}$ as the device is initially accessed by that user. The associated biochemical challenge is to identify the optimal sets of analytes, $x_{n=1,\ldots,N}$, as well as devise biochemical processing to yield suitable outputs $y_{s=1,\ldots,S}$.

We point out that the simplest approach of having $S = N$ and converting each analog chemical data value $x_j$ to analog output signal $y_j$, might not be adequate. Indeed, biochemical data of physiological origins are known to have a significant level of random noise. In most situations where such data are the inputs, few- to multi-input biocatalytic and other chemical processes (as well as processes that yield optical or electronic outputs) have to be utilized to "filter away" the noise and produce "digital" (limited to a couple of specific values or more realistically to narrow ranges of values) outputs. Thus, input analytes are "consolidated" into $S < N$ groups, Equation (1), to yield "digital" outputs:

$$x_{1,\ldots,n_1} \to y_1, \quad x_{n_1,\ldots,n_2} \to y_2, \quad \ldots, \quad x_{n_{S-1},\ldots,n_S} \to y_S. \qquad (1)$$

Substantial work designing such multi-input processes to experimentally realize and theoretically optimize their setup for high-quality "digitized" outputs[12-17] has been carried out in the field of biomolecular computing. This is accomplished by developing biocatalytic (typically, enzymatic) assays supplemented with additional biocatalytic or simple chemical-reaction processes, the latter involving either inputs $x_j$ or outputs $y_j$, which act as "filters" to practically eliminate analog noise and variation in the signals.[12-29]

In a more general context, the concept of multi-(bio)marker monitoring has originally emerged in the medical area, specifically, point-of-care techniques.[30] Physiological (body) fluids such as blood, sweat, saliva, etc. can also be exploited to extract biomarkers that can be utilized for forensic identification. The latter area is rather recent, with novel approaches being developed that implement "combinatorial" monitoring of two biomarkers commonly present in blood, in order to identify characteristics such as sex and ethnicity of a blood sample originator.[7,8] Another combination of two markers, with different temporal profiles, was developed for the determination of the time elapsed since the deposition and the originator's age, of a blood sample found at a crime scene.[9,10]

Specifically for sweat, biochemical assays have been developed capable of determining biological sex from samples obtained from fingertips.[6,11] Some general overviews are available.[31-33] Body fluids can be exploited to reveal information about a particular individual not only via genetic material, but also proteins, low molecular weight compounds, and varying amounts of metabolites that are produced by the body as a result of multiple processes related to metabolism. Metabolism, a process that is regulated by a combination of hormone-based controls[34] is indicative of personal characteristics including, but not limited to, biological sex, age, ethnicity, or health status.

Wearable and mobile electronics, as well as most other input peripherals (keyboards, touch screens) are in permanent or frequent contact with skin, the top layer of which — the epidermis — contains glands that are constantly secreting sweat. The concentrations of various components in sweat are controlled by a complex set of reactions regulated by hormones.[34] Due to the fact that hormone levels significantly vary as a result of age, sex, ethnicity, and lifestyle (e.g., diet and fitness), it is concluded that no two individuals have the same hormone levels as functions of time.[35,36] Thus, the time-series of concentrations of chemical components in an individual's sweat should be specific to that individual. We expect that, the time-dependent concentrations of suitably chosen subsets of the 23 amino acids[37] found in human sweat should suffice to differentiate between individuals.

Here we address the differentiation (authentication) that relies on the analytes in sweat. However, we comment that certain other aspects of the technology are rapidly becoming available.[38] Specifically, not only are sensors undergoing rapid miniaturization,[39] but a wearable wristband technology that electronically detects certain chemicals in sweat has already been demonstrated[39,40] in a different context, and other platforms are being investigated as well.

## 2. AMINO ACID ASSAYS

### 2.1. Bioanalytical Studies and Instrumentation

In order to study the amino acid assays potentially suitable for such goals, gas chromatography (GC) coupled with mass spectrometry (MS) have been utilized for pre-screening. Measurements are performed on an Agilent 5977E GC-MS equipped with ChemStation and MassHunter data analysis software. Given that GC-MS requires volatile components, the amino acids in the samples extracted from sweat are derivatized using MSTFA (N–methyl–N–(trimethylsilyl)trifluoroacetamide) and acetonitrile according to an established procedure described by Thermo Fisher Scientific. [41] In addition, the following protocol [42] is used for analysis of the derivatized amino acid samples: isothermal at 150 °C for 2 minutes, then 150 to 250 °C at a rate of 8 °C/minute, followed by 250 to 310 °C at a rate of 6 °C/minute, and ending with isothermal at 310 °C for 10 minutes. The samples are injected in the splitless mode with a solvent delay of 4 minutes and helium is used as the carrier gas with a constant flow of 1 mL/minute. The injector temperature is held at 250 °C. For MS detection, the masses of the derivatized amino acids are scanned in the quadrupole from m/z 10 to 420.

Such experiments reveal the presence of various amino acids in the sample, as well as their corresponding concentrations. The characterization of the samples from several originators is crucial to support data analysis and, specifically, seek correlations that allow active tracking of life-style for various groups of individuals and ultimately, with a large enough pool of samples, for specific individuals. The results also provide information for designing new bioassays targeting a single analyte or a pool of analytes and their time dependence related to a specific trait or behavioral habits. The concentration distributions for the same individual or a group of similar-trait individuals under similar conditions can also be studied.

## 2.2. Bioaffinity System for Amino Acid Tracking

Various endogenous and exogenous compounds are present in the composition of sweat; amino acids are one of the main components.[43-46] Here, for illustration we describe assays for targeting (converting to a measurable signal) glutamate (Glu), alanine (Ala), and phenylalanine (Phe). Experimental results for the first two of these amino acids are reported in Section 4.2. Glutamate can be analyzed using an enzyme system containing glutamate dehydrogenase (GlDH; E.C 1.4.1.3). GlDH will consume Glu and $NAD^+$ (β-nicotinamide adenine dinucleotide) as substrates, to produce the corresponding amounts of 2-oxoglutarate (KTG), ammonia ($NH_3$), and NADH. As shown in Figure 1, amino acids (glutamate, in this case) can be analyzed in two ways. Pathway A utilizes the UV properties of NADH alone, which can be observed[47] at 340 nm. Pathway B implements phenazine methosulfate (PMS) and Nitroblue tetrazolium (NBT). Here, PMS mediates the reaction of NADH reducing NBT to a colored product, formazan, which can be measured spectrophotometrically at 580 nm.

Changing the starting enzyme and other chemicals as appropriate, the system can be adapted for the detection of alanine and phenylalanine using the same method. Alanine can be analyzed using an enzyme system containing alanine dehydrogenase (AlaDH; E. C. 1.4.1.1). AlaDH consumes alanine and NAD$^+$ to produce pyruvate (Pyr), NH$_3$, and NADH. Phenylalanine dehydrogenase (PheDH; E. C. 1.4.1.20) produces phenylpyruvate (PhPyr), NH$_3$, and NADH in the presence of phenylalanine and NAD$^+$. These two systems can be devised similarly to the glutamate cascade, with either pathway A, direct observation of NADH at 340 nm,[47] or pathway B, conversion of NADH to formazan via NBT/PMS which is observable at 580 nm.[48]

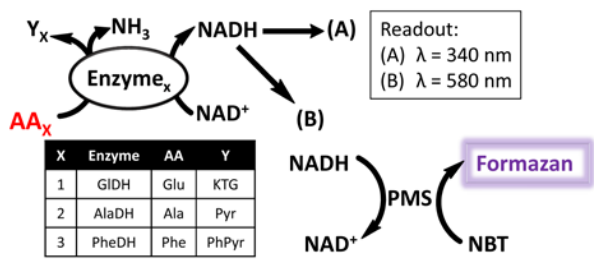

**Figure 1.** Biocatalytic cascade for optical detection of three amino acids, using two detection pathways: (A) measurement of NADH at 340 nm, and (B) conversion of NADH to a visible color (formazan) via NBT/PMS which is observable at 580 nm.

## 3. DATA COLLECTION AND ANALYSIS

In the nomenclature of Equation (1), each assay described in the Sec. 2.2 corresponds to the conversion of a single group of chemical inputs into an output. Bioaffinity-based cascades can be tested and optimized for such assays using the buffer-based samples prepared to mimic the amino acid distributions known to be present in sweat, but ultimately authentic sweat samples should be used. For the successful completion of this phase, we estimate[6-9,11] (from prior experience with forensic assays) that the population used must contain a minimum of 25 volunteers per characteristic being studied. Statistical analysis: Receiver Operating Characteristic (ROC) analysis[49,50] and Area Under the Curve (AUC) analysis[51] have been performed on such results.

Presently, such analyses have been reported to distinguish group characteristics: biological sex, ethnicity, etc.[6-11] for forensics. Scaling this approach up to the authentication of individuals should be possible with the use of time-series of data sets (of $S$ outputs), instead of single-sample (one time) data. As already mentioned, this will likely require the use of several-input processes for each output, Equation (1), to minimize the analog noise in the typical physiological data and produce "digitized" (to narrow output ranges) signals by techniques developed for biomolecular computing.[8-28]

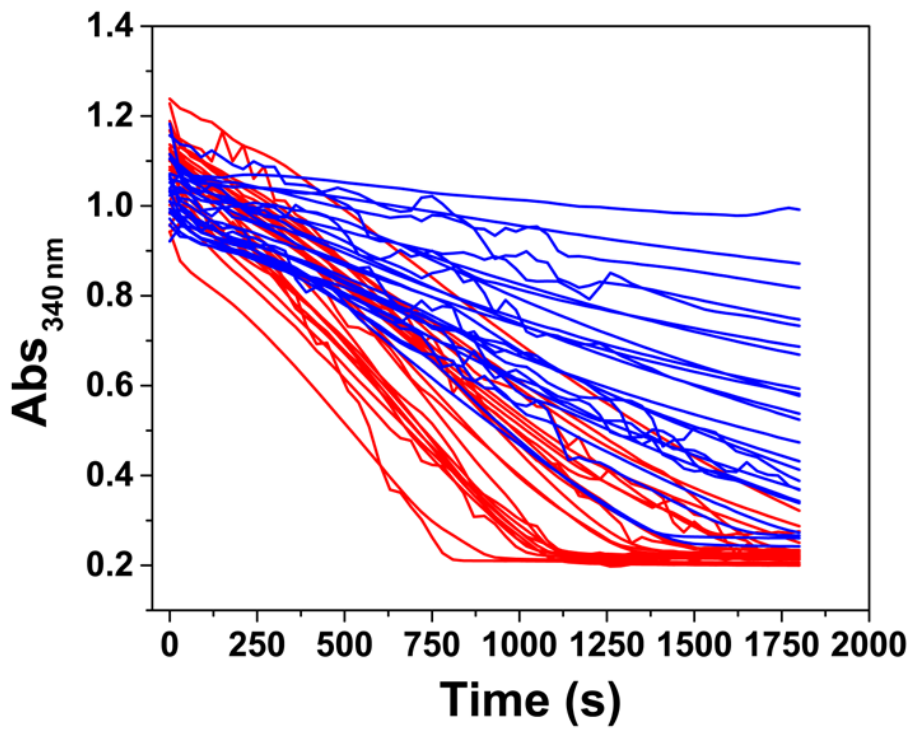

**Figure 2.** Absorbance decrease ($\lambda$ = 340 nm) corresponding to the conversion of NADH to NAD$^+$ via the ALT/LDH assay (see Section 3) using mimicked sweat samples. The red traces indicate female samples; the blue traces indicate male samples.

A large amount of work on "digitalization" in signal processing of noisy data gathered by measurements of signals resulting from multi-step biomolecular, notably, enzyme-catalyzed processes has been carried out.[8-29] Indeed, data in the present case are rather noisy, as exemplified in Figure 2. Once the results of biochemical preprocessing yield the physical and ultimately electronic signals obtained as/from the time series of the outputs $y_{s=1,\ldots,S}(t_0 + k\tau + t_g)$, statistical analysis for determining the feasibility of the authentication mechanism at the level of an individual can be carried out by techniques for time-series/trace-history analysis.[2-5] Continuous monitoring/trace-history recording and analysis of the person's habits and routine behavior is required for active authentication approach.

The approaches considered here can also offer user-group (rather than individual) authentication, e.g., providing age-sensitive access restrictions, gender- and demographic-sensitive marketing, etc. These applications likely require less selective sets of assays (smaller $S$) than individual authentication. We also comment that continuously or regularly (but unobtrusively) collected data can be fed into a fully software-based neural-network type system[5] that can then be trained to follow the user's identity continuously.

## 4. ILLUSTRATIVE RESULTS

### 4.1. ALT/LDH System

The biocatalytic system (cascade of processes) shown in Figure 3, illustrates detection, with results shown in Figure 2, of a single amino acid. Note that in Figure 2 and below, the shown measurement times are the probe-times, $\tau$, discussed in connection with Equation (1).

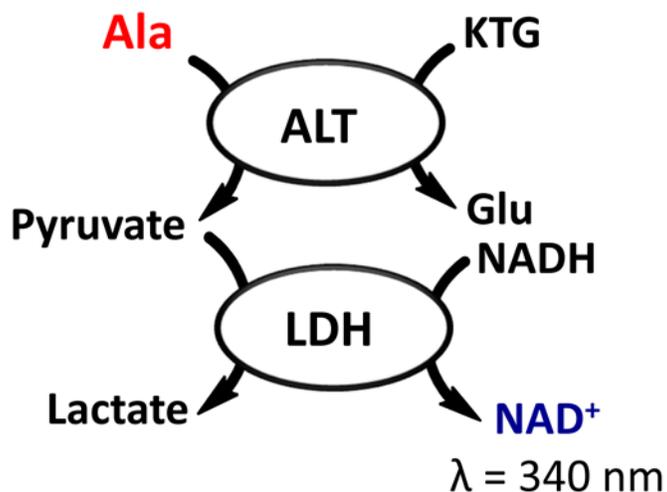

**Figure 3.** The *ALT/LDH* biocatalytic cascade devised for determination of alanine content in sweat.

These and several other measurements (not reported here) were for "mimicked" samples (measurements with authentic sweat samples are exemplified in Section 4.2). The parameters for the amino acid concentration distribution were taken from eccrine sweat and utilized to generate 25 different concentrations for each of the 23 amino acids previously determined to be in sweat from females (and another similar set for males), to give a total of 575 amino acid concentrations (25 for each amino acid). These 25 concentrations capture a plausible range of amino acid concentrations that can be present in sweat. The concentrations were then randomly grouped to create 25 samples (separately for males and females) containing varying quantities of all 23 amino acids. These sets of mimicked sweat samples were prepared in Tris-HCl buffer (pH 7.5) and analyzed using the newly designed and optimized dual-enzyme biocatalytic assay that follows only one specific amino acid, alanine.

This single analyte biocatalytic assay, Figure 3, is based on the combined reactions of two enzymes: alanine transaminase (ALT; E. C. 2.6.1.2.) and lactate dehydrogenase (LDH; E.C. 1.1.1.27). The catalytic process converts alanine and KTG into pyruvate and glutamate, respectively, using ALT. The formed pyruvate then undergoes a reaction that is biocatalyzed by LDH to convert the optically active NADH to $NAD^+$. The NADH generates an optical signal at 340 nm, which decreases as LDH converts[7,8] this compound into $NAD^+$.

The robustness of this assay was assessed in a different context, for detection of traumatic injury markers and even for detection of real injuries in a pig model system.[52] The specificity of the ALT enzyme guarantees that the cascade will only respond to the level of alanine, without any significant interference from the other amino acids present in the samples. The results for this single-analyte system show that the monitoring of one particular amino acid can be achieved even when its concentrations are minute.

**4.2. ALT/POx/HRP System**

A somewhat more complicated cascade than that in the preceding subsection, depicted in Figure 4, demonstrates an unambiguous biological sex identification on the time scale of a couple of minutes. This cascade involves the detection of alanine via the concerted action of three enzymes: ALT, Pyruvate Oxidase (POx, E.C. 1.2.3.3), and Horseradish Peroxidase (HRP, 1.11.1.7). ALT converts L-alanine to pyruvate, which is then subsequently consumed by POx to produce $H_2O_2$. $H_2O_2$ then drives the oxidation of ABTS (2,2'-azino-bis(3-ethylbenzothiazoline-6-sulphonic acid)) via HRP to provide a visible (optical) signal.

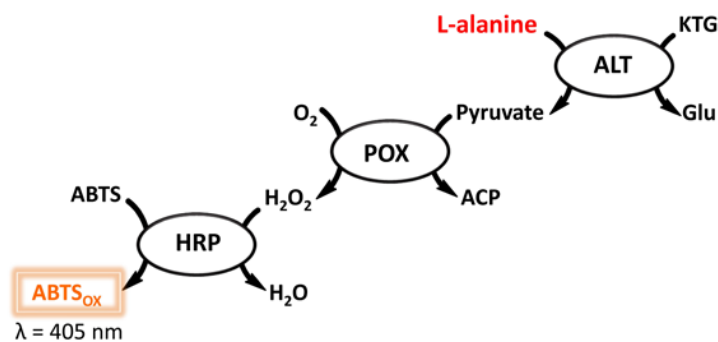

**Figure 4.** The ALT/POx/HRP cascade.

This system has been utilized for the analysis of authentic sweat samples. The success of detecting amino acids in the samples, as well as this single-analyte bioaffinity assay's performance, are demonstrated in Figure 5. The results are definitive for the determination of biological sex, but it can be seen that there is also a discernable difference in signal between different individuals. However, as emphasized in the earlier discussion, multi-analyte analysis, as well as a time-series of measurements for continuous monitoring will be required for a definitive identification of individuals.

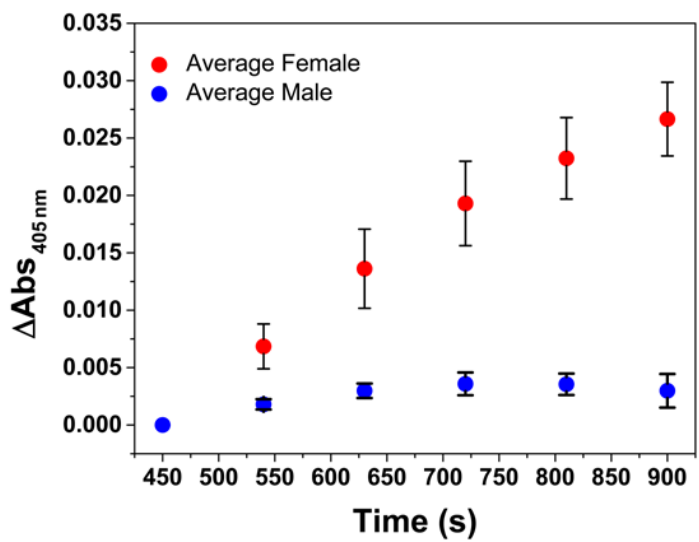

**Figure 5.** Data obtained from authentic sweat samples from males and females, showing averages and spread (as error bars).

### 4.3. GlDH System

Here we illustrate the detection of another amino acid, glutamate, to demonstrate that clear differences between individuals can be observed on time scales, $\tau$, of two minutes and potentially less than that, as shown in Figure 6. The schematic of this system is shown in Figure 7. Glutamate dehydrogenase (GlDH; E.C 1.4.1.3) consumes glutamate and $NAD^+$ as substrates, in order to produce corresponding amounts of KTG, $NH_3$, and NADH. As shown in Figure 7, glutamate can be analyzed in three ways. Pathways A and B were already described in connection with the Figure 1. In Pathway C, $NADH_{ox}$ produces $H_2O_2$ in the presence of NADH and oxygen. $H_2O_2$ is then consumed by another enzyme, HRP, in order to generate a luminescent response at about 425-445 nm via Luminol.[53]

Note that the latter pathway yields $H_2O_2$, which is directly convertible to electronic signals,[54,55] as frequently used in sensor devices. While optical characterization techniques are

– 13 –

convenient for basic-science studies, conversion to electrical signals may be advantageous in applications.

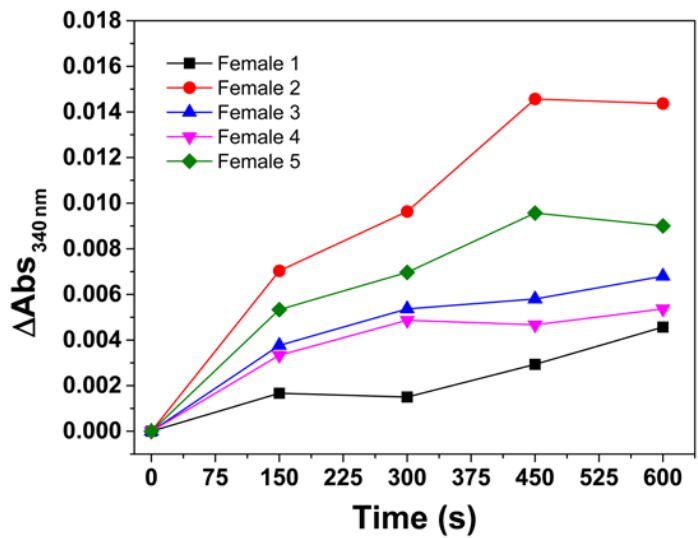

**Figure 6.** Results for the GlDH assay using authentic female sweat samples obtained from fingertips.

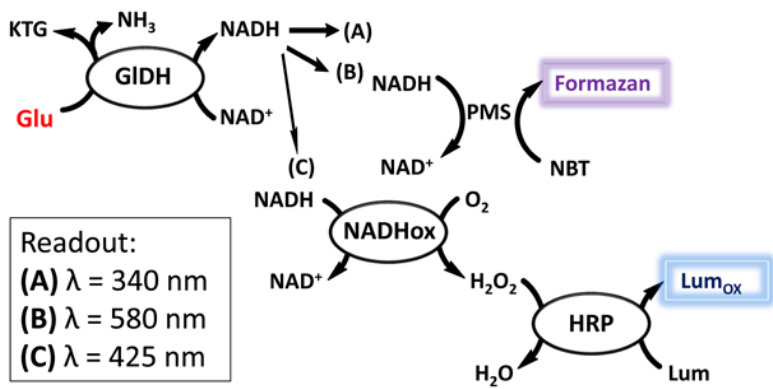

**Figure 7.** The biocatalytic cascade used for the detection of glutamate, along with the three detection pathways: (A) direct measurement of NADH at 340 nm, (B) conversion of NADH to a visible color (formazan) via NBT/PMS which is observable at 580 nm, and (C) oxidation of NADH to produce hydrogen peroxide which is further used to oxidize Luminol via HRP to produce bioluminescence at 425 nm.

## 5. MULTI-INPUT SYSTEMS

As explained earlier, it may be advantageous to use several-input processes before converting the output into a physical and ultimately electronic form. Here we outline the design of such cascades for combinations of three amino acids, alanine (Ala), glutamate (Glu), and aspartate (Asp), two of which were encountered in the illustrative single-input studies reported in Section 4.

We first consider, Figure 8, the two-input Ala/Glu system that utilizes the enzyme ALT to convert alanine to pyruvate (Pyr) which then drives the consumption of KTG as the second substrate for this enzyme, to produce glutamate. The amount of glutamate produced with ALT is added to the glutamate that is present in the sample as an input. The enzyme glutamate oxidase (GlOx) consumes the Glu to drive the production of $H_2O_2$. $H_2O_2$ can be converted to electronic signal, as outlined in Section 6 below, or it can be consumed by the enzyme HRP, Figure 8, to drive the oxidation of ABTS which provides a visible signal[56] at the wavelength of 405 nm. This system effectively adds the amounts of the two inputs to produce one output, so it is not of the "digital" nature. However, if needed it can be made "digital" in its response, by adding additional "filtering" chemical or biocatalytic processes.[12-29]

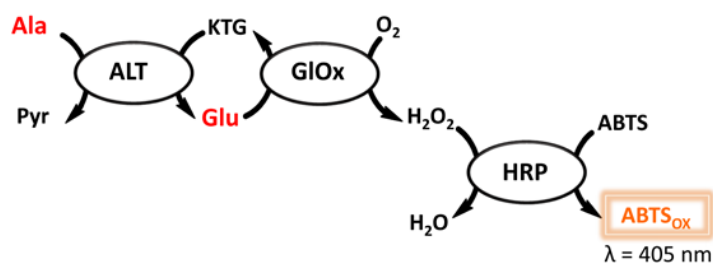

**Figure 8.** Ala/Glu Assay.

Let us next consider a design for a two-input Asp/Glu system, where Asp denotes aspartate. This system is similar to the Ala/Glu system just considered, but here, see Figure 9, aspartate transaminase (AST) is used for targeting of aspartate. In Figure 9 and below, OAC denotes oxaloacetate.

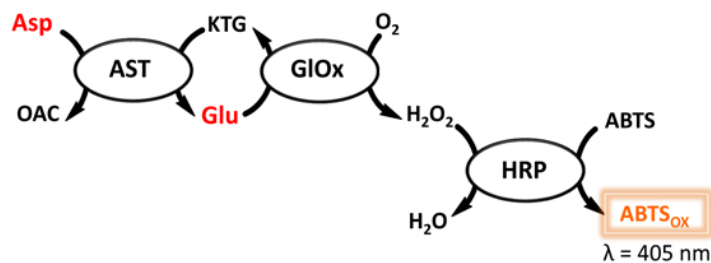

**Figure 9.** Asp/Glu Assay.

A combination of the two abovementioned systems is possible, to yield a three-input biocatalytic cascade, as shown in Figure 10, involving four enzymes. Alternative versions of these systems are also possible, involving for instance use of enzyme GlDH, with which Glu would be consumed to drive the production of NADH instead of $H_2O_2$. This does not provide a visible signal, but does produce a readable signal in the ultraviolet range (340 nm).

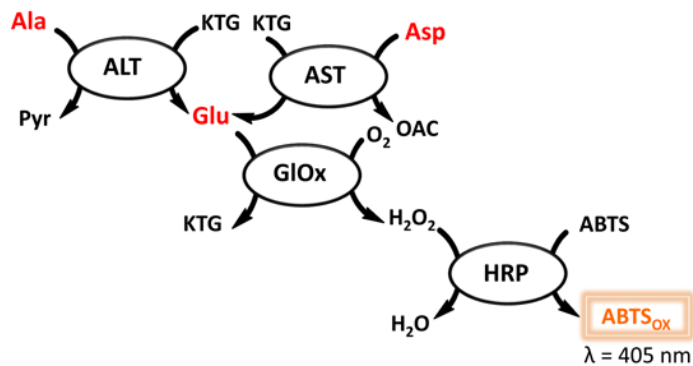

**Figure 10.** Ala/Asp/Glu Assay.

## 6. DISCUSSION AND SUMMARY

The chemical signals obtained with processes described in the preceding sections, some already demonstrated, as well as other multi-input systems that require not only future realizations but also "optimization" for digital outputs, can be used directly, converted to optical (absorbance, as in Figures 4-6, or fluorescence) outputs for studies of time-series of signal values obtained. However, ultimately, for devices the priority should be given to systems that also allow output in the form of electronic signals. In the considered cascades, the designs that yield $H_2O_2$ as the output are thus preferable. Indeed, consumption of $H_2O_2$ biocatalyzed by the enzyme HRP can yield electronic signals.

An electrode platform can be based on commercial screen-printing technology. Au-SPEs (screen-printed gold electrodes) can be utilized in small volume cells for optimization of the assays and their performance. The electrodes are modified with the necessary enzymatic machinery for the specific detection cascades in order to retain enzymatic activity, increase the electroactive area and diminish undesired adsorption.[57,58] This concept utilizes the detection of $H_2O_2$ by biocatalytic action of HRP directly on the electrode surface, allowing for quantitative detection of $H_2O_2$.

Thus, the considered technology is suitable for the future development of actual chips (sensors) and wearable devices. Research in this area has already demonstrated that human sweat can be analyzed using a wearable device for continuous monitoring of an individual's state of health.[40]

In summary, we reviewed novel emerging concepts in utilizing biocatalytic cascades for detection of amino acids in sweat, with promise of applications for active authentication and continuous device-user tracking for cybersecurity of mobiles, wearable, and similar devices. Our emphasis has been on the biochemistry aspects of this approach, with some of the considered systems already demonstrated in the preliminary studies, whereas other, more complicated systems are still in design stages.

**GRAPHICAL ABSTRACT AND CAPTION FOR TABLE OF CONTENTS**

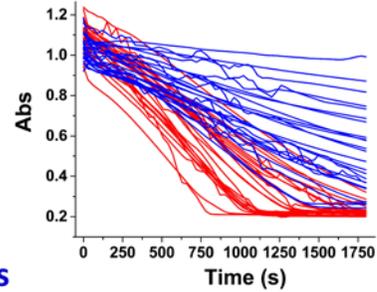
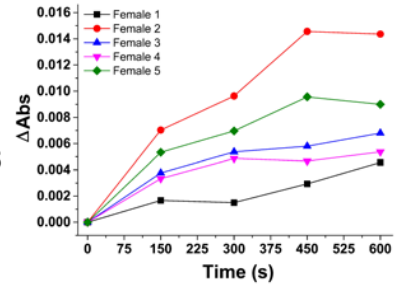

**Biometric-based cybersecurity** systems for active authentication by continuous tracking utilizing biochemical processing of metabolites present in skin secretions are described. We argue that amino acids found in sweat can be exploited for the establishment of an amino acid profile capable of identifying an individual user of a mobile or wearable device. Individual and combinations of amino acids processed by biocatalytic cascades yield physical (optical or electronic) signals, providing a time-series of several outputs that, in their entirety, should suffice to authenticate a specific user based on standard statistical criteria.

# REFERENCES

[1] T. Matsumoto, H. Matsumoto, K. Yamada, S. Hoshino, *Proc. SPIE* **2002**, *4677*, 275–289; DOI 10.1117/12.462719

[2] U. Mahbub, R. Chellappa, PATH: Person Authentication using Trace Histories, *Proc. 7th IEEE Annual Ubiquitous Computing, Electronics & Mobile Communication Conference* **2016**, New York, USA; DOI 10.1109/UEMCON.2016.7777911

[3] C. McCool, S. Marcel, A. Hadid, M. Pietikainen, P. Matejka, J. Cernocky, N. Poh, J. Kittler, A. Larcher, C. Levy, D. Matrouf, J.-F. Bonastre, P. Tresadern, T. Cootes, Bi-modal person recognition on a mobile phone: using mobile phone data, *Proc. IEEE International Conference on Multimedia and Expo Workshops* **2012**, Melbourne, Australia; DOI 10.1109/ICMEW.2012.116

[4] U. Mahbub, S. Sarkar, V. M. Patel, R. Chellappa, Active User Authentication for Smartphones: A Challenge Data Set and Benchmark Results, *Proc. 8th IEEE International Conference on Biometrics: Theory, Applications, and Systems* **2016**, Buffalo, USA; DOI 10.1109/BTAS.2016.7791155

[5] C.-K. Chen, C.-L. Lin, Y.-M. Chiu, Individual Identification Based on Chaotic Electrocardiogram Signals, *Proc. 6th IEEE Conference on Industrial Electronics and Applications* **2011**, Beijing, China; DOI 10.1109/ICIEA.2011.5975879

[6] E. Brunelle, C. Huynh, A. M. Le, L. Halámková, J. Agudelo, J. Halámek, *Anal. Chem.* **2016**, *88*, 2413–2420; DOI 10.1021/acs.analchem.5b03323

[7] S. Bakshi, L. Halámková, J. Halámek, E. Katz, *Analyst* **2013**, *139*, 559–563; DOI 10.1039/C3AN02055J

[8] F. Kramer, L. Halámková, A. Poghossian, M. J. Schöning, E. Katz, J. Halámek, *Analyst* **2013**, *138*, 6251–6257; DOI 10.1039/C3AN01062G

[9] J. Agudelo, C. Huynh, J. Halamek, *Analyst* **2014,** *140*, 1411–1415; DOI 10.1039/C4AN02269F

[10] J. Agudelo, L. Halámková, E. Brunelle, R. Rodrigues, C. Huynh, J. Halámek, *Anal. Chem.* **2016,** *88*, 6479–6484; DOI 10.1021/acs.analchem.6b01169

[11] C. Huynh, E. Brunelle, L. Halámková, J. Agudelo, J. Halámek, *Anal. Chem.* **2015**, *87*, 11531–11536; DOI 10.1021/acs.analchem.5b03323

[12] V. Privman, O. Zavalov, L. Halamkova, F. Moseley, J. Halamek, E. Katz, *J. Phys. Chem. B* **2013**, *117*, 14928–14939; DOI 10.1021/jp408973g

[13] S. Bakshi, O. Zavalov, J. Halámek, V. Privman, E. Katz, *J. Phys. Chem. B* **2013**, *117*, 9857–9865; DOI 10.1021/jp4058675

[14] J. Halámek, O. Zavalov, L. Halámková, S. Korkmaz, V. Privman, E. Katz, *J. Phys. Chem. B* **2012**, *116*, 4457–4464; DOI 10.1021/jp300447w

[15] J. Halamek, J. Zhou, L. Halamkova, V. Bocharova, V. Privman, J. Wang, E. Katz, *Anal. Chem.* **2011**, *83*, 8383–8386; DOI 10.1021/ac202139m

[16] V. Privman, J. Halamek, M. Arugula, A., D. Melnikov, V. Bocharova, E. Katz, *J. Phys. Chem. B* **2010**, *114*, 14103–14109; DOI 10.1021/jp108693m

[17] V. Privman, M. A. Arugula, J. Halámek , M. Pita, E. Katz, *J. Phys. Chem. B* **2009**, *113*, 5301–5310; DOI 10.1021/jp810743w

[18] S. Domanskyi, V. Privman, *J. Phys. Chem. B* **2012**, *116*, 13690–13695; DOI 10.1021/jp309001j

[19] U. Pischel, J. Andreasson, D. Gust, V. F. Pais, *ChemPhysChem* **2013,** *14*, 28–46; DOI 10.1002/cphc.201200157

[20] Y. Benenson, *Nature Rev. Genet.* **2012**, *13*, 455–468; DOI 10.1038/nrg3197

[21] M. N. Stojanovic, D. Stefanovic, S. Rudchenko, *Acc. Chem. Res.* **2014**, *48*, 1845–4852; DOI 10.1021/ar5000538

[22] M. N. Stojanovic, D. Stefanovic, *J. Comput. Theor. Nanosci.* **2011**, *8*, 434–440; DOI 10.1166/jctn.2011.1707

[23] Y. Liu, E. Kim, I. M. White, W. E. Bentley, G. F. Payne, *Bioelectrochem.* **2014**, *98*, 94–102; DOI 10.1016/j.bioelechem.2014.03.012

[24] H. Jiang, M. D. Riedel, K. K. Parhi, *IEEE Design Test Computers* **2012**, *29*, 21–31; DOI 10.1109/MDT.2012.2192144

[25] P. Hillenbrand, G. Fritz, U. Gerland, *PLOS ONE* **2013**, *8*, article e68345 (10 pages); DOI 10.1371/journal.pone.0068345

[26] C. G. Bowsher, *J. R. Soc. Interf.* **2011**, *8*, 168–200; DOI 10.1098/rsif.2010.0287

[27] J. Halámek, V. Bocharova, S. Chinnapareddy, J. R. Windmiller, G. Strack, M. C. Chuang, J. Zhou, P. Santhosh, G. V. Ramirez, M. A. Arugula, J. Wang, E. Katz, *Mol. Biosyst.* **2010**, *6*, 2554–2560; DOI 10.1039/c0mb00153h

[28] S. P. Rafael, A. Vallée-Bélisle, E. Fabregas, K. Plaxco, G. Palleschi, F. Ricci, *Anal. Chem.* **2012**, *84*, 1076–1082; DOI 10.1021/ac202701c

[29] A. Vallée-Bélisle, F. Ricci, K. W. Plaxco, *J. Am. Chem. Soc.* **2012,** *134*, 2876–2879; DOI 10.1021/ja209850j

[30] P. B. Luppa, C. Müller, A. Schlichtiger, H. Schlebusch. *Trends Anal. Chem.* **2011**, *30*, 887–898; DOI 10.1016/j.trac.2011.01.019

[31] J. Woolford, A biochemical eyewitness, *Chemistry World*, 5 September **2013**; http://www.chemistryworld.com/news/a-biochemical-eyewitness/6557.article

[32] K. Muirhead, Biomarkers leave gender clues at crime scene, *Chemistry World*, 15 January **2014**; http://www.chemistryworld.com/news/biomarkers-leave-gender-clues-at-crime-scene/6972.article

[33] E. Stoye, Cutting edge chemistry in 2013, *Chemistry World*, 12 December **2013**; http://www.chemistryworld.com/news/cutting-edge-chemistry-in-2013/6892.article

[34] A. J. Thody, S. Shuster, *Physiol. Rev.* **1989**, *69*, 383–416; http://physrev.physiology.org/content/69/2/383

[35] J. W. Finkelstein, H. P. Roffwarg, R. M. Boyar, J. Kream, L. Hellman, *J. Clin. Endocrinol. Metab.* **1972**, *35*, 665–670; DOI 10.1210/jcem-35-5-665

[36] J. C. Pruessner, C. Kirschbaum, G. Meinlschmid, D. H. Hellhammer, *Psychoneuroendocrinol.* **2003**, *28*, 916–931; DOI 10.1016/S0306-4530(02)00108-7

[37] M. Harker, C. R. Harding, *Int. J. Cosmet. Sci.* **2013,** *35*, 163–168; DOI 10.1111/ics.12019

[38] D. Sharon, Are Smart Mini Sensors the Next Big Thing? *Live Science*, 12 August **2015**; http://www.livescience.com/51832-smart-mini-sensors-are-changing-technology.html

[39] M. Nguyen, We Will Make You Sweat, *Wearable Technologies*, 17 February **2016**; http://www.wearable-technologies.com/2016/02/we-will-make-you-sweat

[40] W. Gao, S. Emaminejad, H. Y. Y. Nyein, S. Challa, K. Chen, A. Peck, H. M. Fahad, H. Ota, H. Shiraki, D. Kiriya, D.-H. Lien, G. A. Brooks, R. W. Davis, A. Javey, *Nature* **2016**, *529*, 509–514; DOI 10.1038/nature16521

[41] A. I. Khan, Application Note 20560: GC Analysis of Derivatized Amino Acids, *Thermo Fisher Scientific*, Runcorn, UK; http://static.thermoscientific.com/images/D22161~.pdf M.

[42] S. Khan, G. Thouas, W. Shen, G. Whyte, G. Garnier, *Anal. Chem.* **2010**, *82*, 4158–4164; DOI 10.1021/ac100341n

[43] F. Costa, D. Howes, S. Margen, *Am. J. Clin. Nutr.* **1969**, *22*, 52–58; http://ajcn.nutrition.org/content/22/1/52

[44] S. W. Hier, T. Cornbleet, O. Bergeim, *J. Biol. Chem.* **1946**, *166*, 327–333; http://www.jbc.org/content/166/1/327.full.pdf+html

[45] V. P. Kutyshenko, M. Molchanov, P. Beskaravayny, V. N. Uversky, M. A. Timchenko, *PLOS ONE* **2011**, *6*, article e28824 (9 pages); DOI 10.1371/journal.pone.0028824

[46] N. Liappis, A. Jäkel, *Arch. Dermatol. Res.* **1975**, *254*, 185–203; DOI 10.1007/BF00586893

[47] Determination of NADH Concentrations with the Synergy™ 2 Multi-Detection Microplate Reader using Fluorescence or Absorbance, *BioTek Instruments*, Winooski, USA, **2006**; http://www.biotek.com/resources/docs/NADH_App_Note.pdf

[48] NBT, *Sigma-Aldrich*, St. Louis, USA; http://www.sigmaaldrich.com/catalog/product/roche/11585029001

[49] X. Robin, N. Turck, A. Hainard, N. Tiberti, F. Lisacek, J.-C. Sanchez, M. Müller, *BMC Bioinformatics* **2011**, *12*, article 77 (8 pages); DOI 10.1186/1471-2105-12-77

[50] R Development Core Team, R: A Language and Environment for Statistical Computing, *The R Foundation for Statistical Computing*, Vienna, Austria, **2011**; http://www.r-project.org/

[51] E. R. DeLong, D. M. DeLong, D. L. Clarke-Pearson, *Biometrics* **1988**, *44*, 837–845; DOI 10.2307/2531595

[52] L. Halámková, J. Halámek, V. Bocharova, S. Wolf, K. E. Mulier, G. Beilman, J. Wang, E. Katz, *Analyst* **2012**, *137*, 1768–1770; DOI 10.1039/C2AN00014H

[53] J. Arnhold, S. Mueller, K. Arnold, E. Grimm, *J. Biolumin. Chemilumin.* **1991**, *6*, 189–192; DOI 10.1002/bio.1170060309

[54] A. M. Farah, F. T. Thema, E. D. Dikio, *Int. J. Electrochem. Sci.* **2012**, 7, 5069–5083; http://www.electrochemsci.org/papers/vol7/7065069.pdf

[55] A. L. Sanford, S. W. Morton, K. L. Whitehouse, H. M. Oara, L. Z. Lugo-Morales, J. G. Roberts, L. A. Sombers, *Anal. Chem.* **2010**, *82*, 5205–5210; DOI 10.1021/ac100536s

[56] ABTS™, *Sigma-Aldrich*, St. Louis, USA; http://www.sigmaaldrich.com/catalog/product/roche/10102946001?lang=en®ion=US

[57] F. Wachholz, H. Duwensee, R. Schmidt, M. Zwanzig, J. Gimsa, S. Fiedler, G.-U. Flechsig, *Electroanal.* **2009**, *21*, 2153–2159; DOI 10.1002/elan.200904665

[58] T.-F. Tseng, Y.-L. Yang, M.-C. Chuang, S.-L. Lou, M. Galik, G.-U. Flechsig, J. Wang, *Electrochem. Commun.* **2009**, 11, 1819–1822; DOI 10.1016/j.elecom.2009.07.030

**JOURNAL ISSUE COVER**

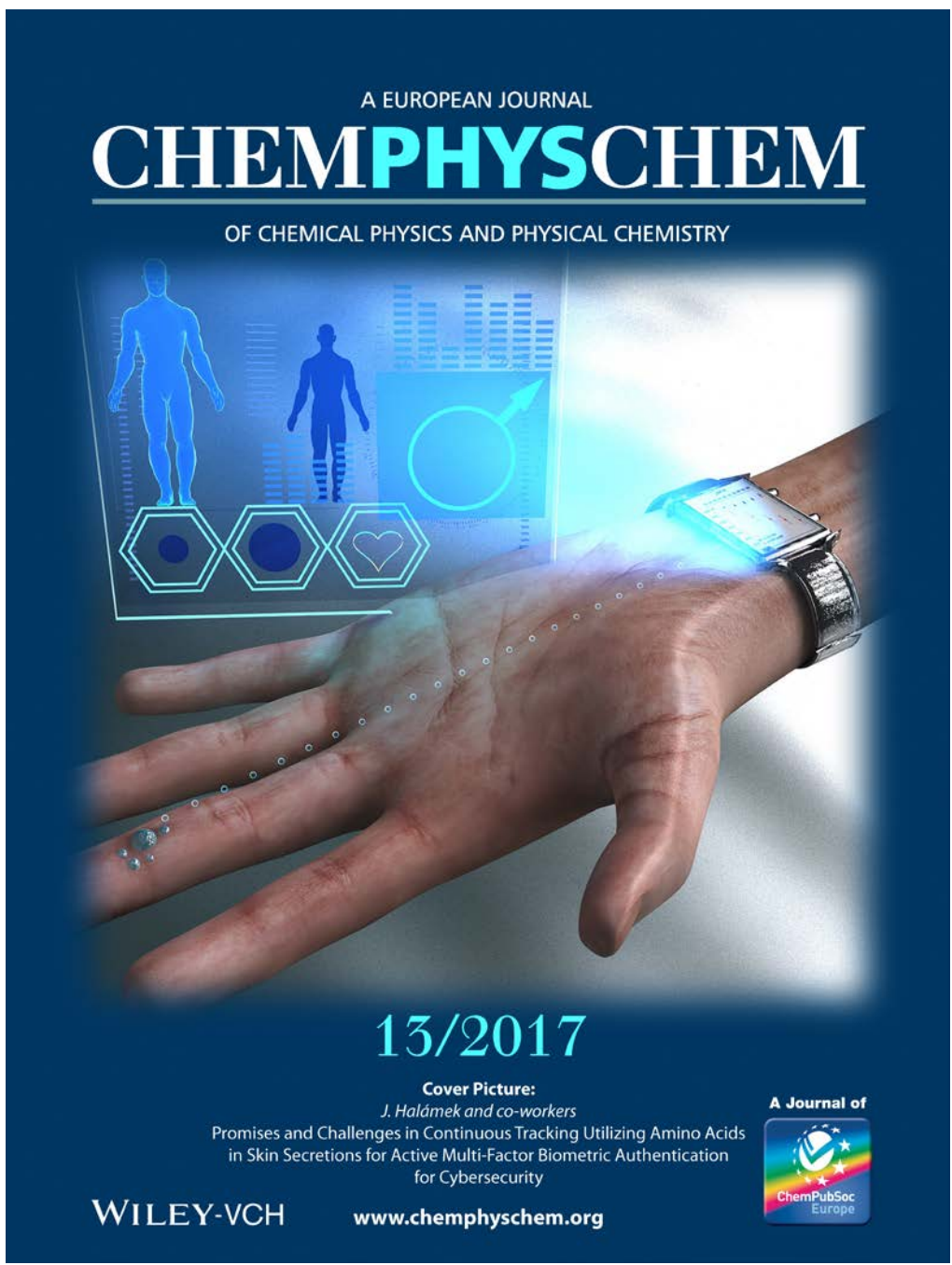